\documentclass[12pt]{article}
\nonstopmode
\usepackage[x11names]{xcolor}
\usepackage{amsmath}
\usepackage[hmargin=1in,vmargin=1in]{geometry}
\usepackage{float}
\usepackage{longtable}
\usepackage{graphicx}
\usepackage{multirow}
\usepackage{chngcntr}
\usepackage{url}

\counterwithin{figure}{section}
\counterwithin{table}{section}

\usepackage{times}

\begin{document}


\title{An Empirical Study on the Effects of the America Invents Act on Patent Applications Owned by Small Businesses}

\author{Yoo Jeong Han\\
  Department of Statistics\\
  University of California, Los Angeles}

\date{February 2021}

\maketitle

\begin{abstract}
  This paper evaluates the heterogenous impacts of the America Invents Act of 2011 (AIA) on patent applications for small and large businesses. Using data collected from the United States Patent and Trademark Office and Google Patents, I compare how the probability of successfully overcoming an initial rejection is affected by the AIA for small- and large-business applicants, respectively. This comparison is achieved by analyzing the data using a difference-in-differences approach. Results suggest that after the enactment of the AIA, small-business applicants were relatively favored when compared against large-business applicants. This effect is statistically significant and also practically large.
\end{abstract}

\newpage
\section{Introduction}

It has been argued that small businesses have been unfavorably affected
by the major reforms (Braun, 2012) of the Leahy-Smith America Invents
Act of 2011 (AIA). Vandenburg (2014) and Sutton (2014) propose that the
transition from the First-To-Invent system to the First-To-File system
(Section 3, AIA) favors large businesses that have the financial
resources to accommodate faster filing, thus placing small businesses at
a relative disadvantage. Case (2013) notes the AIA’s elimination of the
post-disclosure grace period as another source of disadvantage. The lack
of a post-disclosure grace period forces inventors to either file before
fully developing their ideas or to fully develop their
ideas while at risk of premature disclosure by others. This is
advantageous to larger businesses that have relatively abundant
financial resources for faster development.

While Vandenburg (2014), Sutton (2014), and Case (2013) focus on small
businesses that have not yet filed their patent applications, Lerner,
Speen, and Leamon (2015) attend to the possible disadvantages to
small-business patent owners by measuring market reaction to the AIA for
publicly-traded patent-owning small businesses. Their study, however,
does not find clear evidence of unfavorable market reactions, and
reports only minimal differential impact by the AIA.

In this paper, I focus on another group of small businesses that may be affected by the AIA enactment -- those that have applications that are undergoing pre-grant patent prosecution. Specifically, I compare how the probability of successfully overcoming an initial rejection is affected by the AIA for small- and large-business applicants, respectively. 
This comparison pertains to a phase
of patent prosecution called the `Search and Examination phase,'
summarized as follows.


A patent application is examined for its
patentability by a patent examiner at the United States Patent and
Trademark Office (USPTO). If the application is deemed unpatentable as
is, the applicant would most likely receive a Non-Final Rejection. This
Non-Final Rejection details the arguments against patent issuance of the
current application. The applicant’s response to the Non-Final Rejection
attempts to overcome the examiner’s listed arguments by arguing against
the examiner or by amending the current application. The successful
response places the application in a position ready for patent issuance.
The unsuccessful response fails to convince the examiner that the
application is ready for patenting. This can be because the response
inadequately addresses the issues set forth in the Non-Final Rejection
or because the response creates new reasons for rejection in its
amendments.

To identify the AIA’s effect on applications in the early Search and Examination phase,
I examine the success rate, i.e., the proportion of patent applications
that successfully respond to their initial Non-Final Rejections. I
compare the difference in success rate of small-business applicants to
that of large-business applicants before and after the AIA. If the AIA
does not favor or disfavor small businesses in this phase, the
difference in success rate for small-business applicants should roughly
equal that of large-business applicants. In other words, the two
differences would be no different.

For this comparison, I take the difference-in-differences (DID)
approach. In applying the DID method, I assume that trends in the
success rate due to uncontrolled characteristics are independent of
business size. This is called the parallel trends assumption and is key
to the validity of the DID analysis. (See Section 3 for details; see
also Wing, Simon, and Bello-Gomez, 2018, and Schiozer, Mourad, and
Martins, 2021.)

It is notable that when a new law is enacted, the behavioral adjustment of its related entities, including applicants, patent lawyers and agents, and USPTO examiners, may not be immediately clear. This is especially relevant to the research in this paper in that it involves variables that are difficult to measure with precision. Some of these variables are the quality of a small-business applicant’s attempt to overcome a rejection and the examiner’s judgement on that small-business applicant’s attempt. These variables make it difficult to deduce the AIA’s effects on small-business applicants using \emph{a priori} arguments, which in turn, makes the current research topic more interesting. Furthermore, to the best knowledge of the author, this paper is the first attempt in current literature to explain the AIA’s effects on patent applications owned by small and large businesses that are undergoing pre-grant prosecution.

The results of this paper show that there is a significant difference in
 differences. Small business applicants experienced more success in
their Non-Final Rejection responses than large-business applicants, at
least for those which were ultimately issued as patents.

The rest of this paper is organized as follows. Section \ref{sec:data}
describes data sources and the measurement of variables and provides
summary statistics. A brief summary of the DID methodology and
estimation results are presented in Section \ref{sec:analysis}. Section
\ref{sec:conclude} contains concluding remarks.

\section{Data}
\label{sec:data}

\subsection{Data sources and measurements}

The data is collected through a series of maneuvers within the USPTO
website and Google Patents. I first used the USPTO’s PatFT search engine to retrieve a
comprehensive list of 1,866 patents that have an application date
ranging from the 2nd to the 6th of January 2010. I obtained the patent
number and application number for each of the patents on this list. The
patent number’s format contains information regarding its invention
subject category, i.e., whether the patent is a utility, design, or
plant patent. Using this information, I exclusively retained the numbers
for utility patents. I then used each of the retained application
numbers to locate and scrape the corresponding utility patent’s public
data on the USPTO’s electronic filing and patent application management
tool, Patent Center. Patent Center provides detailed information for
each individual patent such as the history of documents and transactions
between the applicant and the USPTO from application filing to patent
issuance. It also provides information regarding the patent’s
application meta data, continuity, patent term adjustment, and attorney
information and more. For example, Patent Center provides that Patent
No. 7,992,975 (Non-Conductive Fluid Droplet Forming Apparatus and
Method) falls under Class 347 and Group Art Unit 2861, has business
entity status LARGE/UNDISCOUNTED, has a particular correspondence
address in the United States, has four attorneys, claims priority to two
parent applications (1 DIV type, 1 PRO type), has a Priority Date of
10/04/2004, has two inventors (all US inventors), and has 44 incoming
and outgoing documents from application filing to patent issuance. After
filing the application on 01/05/2010, the first USPTO Office Action is a
Non-Final Rejection (CTNF), which took place on 02/18/2011. This is
followed by the applicant’s response to the Non-Final Rejection (REM) on
04/07/2011, and a Notice of Allowance (NOA) on 04/15/2011. Using the
data scraped from Patent Center, I narrowed the list of utility patents
down to 1,467, which excludes national-stage applications originating
from foreign or international patent applications.

An Office Action is defined in this paper as a Notice of Allowance
(NOA), Ex Parte Quayle Action (CTEQ), Requirement for
Restriction/Election (CTRS), Non-Final Rejection (CTNF), or Final
Rejection (CTFR). Using the scraped data from Patent Center, I identify
the first Office Action (OA1) the applicant receives for each utility
patent. If OA1 is a Non-Final Rejection, I manually downloaded the
OA1 document in XML or PDF form. This part of the data collection
process could not be automated because Patent Center does not allow
access to the documents in programmable ways that I know of. For each
OA1 that is a Non-Final Rejection, I extracted the number of claims that
are rejected under each of 35 USC § 112, 101, 102(a), 102(b), 102(e),
103(a) and 103(e). This process was automated for the vast majority of
these files in XML format, while I manually counted the number of claims
rejected under each section for the remaining few in PDF format.

Finally, I obtained the number of independent claims for each utility patent. This information was gathered by scraping the text of each claim in each utility patent from Google Patents. The text of each claim was processed to determine whether that claim is independent or dependent.

\subsection{Data summary}

This section summarizes information in the data set relevant to the
research question. The types of first Office Actions (OA1) are Notice of
Allowance (NOA), Ex Parte Quayle Action (CTEQ), Requirement for
Restriction/Election (CTRS), Non-Final Rejection (CTNF), and Final
Rejection (CTFR). The distribution of OA1 types is summarized as
follows.

\begin{table}[H]
  \caption{Distribution of OA1 types}
  \label{tbl:oa1}
  \begin{center}
    \begin{tabular}{|c|c|c|c|c|c|c|}
      \hline
      Type & NOA & CTEQ & CTRS & CTNF & CTFR & Total\\ \hline
      Number & 176 & 21 & 276 & 991 & 3 & 1,467\\ \hline
      Share (\%) & 12.0 & 1.4 & 18.8 & 67.6 & 0.2 & 100.0\\ \hline
    \end{tabular}
  \end{center}
\end{table}

Out of 1,467 patents, 176 applications (12.0\%) were allowed in OA1. OA1 is a CTNF type in 991 applications (67.6\%), and the other 300 applications (20.4\%) had the remaining types.

For the 991 applications that received a CTNF-type OA1, the types for
the second Office Action (OA2) are distributed as in Table
\ref{tbl:oa2}. Among the 991 applications which received CTNF-type OA1s,
363 of them have a CTFR-type (36.6\%), 78 of them have a CTNF-type
(7.9\%), and 539 of them have a NOA-type (54.4\%) OA2.

\begin{table}[H]
  \caption{Distribution of OA2 types after CTNF-type OA1}
  \label{tbl:oa2}
  \begin{center}
    \begin{tabular}{|c|c|c|c|c|c|c|}
      \hline
      Type & NOA & CTEQ & CTRS & CTNF & CTFR & Total\\ \hline
      Number & 539 & 2 & 9 & 78 & 363 & 991\\ \hline
      Share (\%) & 54.4 & 0.2 & 0.9 & 7.9 & 36.6 & 100.0\\ \hline
    \end{tabular}
  \end{center}
\end{table}

The composition of the business entity status is presented in Table \ref{tbl:bsize0} for the total 1,467 applications.

\begin{table}[H]
  \caption{Distribution of business entity status category (all patents)}
  \label{tbl:bsize0}
  \begin{center}
    \begin{tabular}{|c|c|c|c|c|}
      \hline
      Type & Large & Small & Micro & Total\\ \hline
      Number & 1,138 & 305 & 24 & 1,467\\ \hline
      Share (\%) & 77.6 & 20.8 & 1.6 & 100.0\\ \hline
    \end{tabular}
  \end{center}
\end{table}

A vast majority of the patent applications have applicants with business
entity status `LARGE/ UNDISCOUNTED', while the proportion of those with
`SMALL' or `MICRO' status is 22.4\%. Of those with a CTNF-type OA1, the
distribution is as given in Table \ref{tbl:bsize1}. The proportion of
small businesses (including SMALL and MICRO) is slightly smaller
(21.3\%) when conditioned on the event of a CTNF-type OA1.

\begin{table}[H]
  \caption{Distribution of business entity status for applications with CTNF-type OA1}
  \label{tbl:bsize1}
  \begin{center}
    \begin{tabular}{|c|c|c|c|c|}
      \hline
      Type & Large & Small & Micro & Total\\ \hline
      Number & 780 & 194 & 17 & 991\\ \hline
      Share (\%) & 78.7 & 19.6 & 1.7 & 100.0\\ \hline
    \end{tabular}
  \end{center}
\end{table}

A full description of the OA1 types for each business entity status is given in Table \ref{tbl:xtable.oa1}.

\begin{table}[H]
  \caption{Cross table for OA1 types}
  \label{tbl:xtable.oa1}
  \begin{center}
    \begin{tabular}{|c|c|c|c|c|c|c|}
      \hline
      Type & NOA & CTEQ & CTRS & CTNF & CTFR & Sum\\ \hline
      \multirow{2}{*}{Large} & 147 & 17 & 191 & 780 & 3 & 1,138\\
      & (12.9) & (1.5) & (16.8) & (68.5) & (0.3) & (100.0)\\ \hline
      \multirow{2}{*}{Small} & 27 & 4 & 80 & 194 & 0 & 305\\
      & (8.9) & (1.3) & (26.2) & (63.6) & (0.0) & (100.0)\\ \hline
      \multirow{2}{*}{Micro} & 2 & 0 & 5 & 17 & 0 & 24\\
      & (8.3) & (0.0) & (20.8) & (70.8) & (0.0) & (100.0)\\ \hline
      \multirow{2}{*}{Sum} & 176 & 21 & 276 & 991 & 3 & 1,467\\
      & (12.0) & (1.4) & (18.8) & (67.6) & (0.2) & (100.0)\\ \hline
    \end{tabular}
  \end{center}
  \footnotesize
  \emph{Note:} Row-wise percentage shares are in parentheses.
\end{table}

The proportion of NOA-type OA1 is slightly higher for large businesses in comparison to small (and micro) ones, and this difference turns out to be statistically significant at the 5\% level. The proportion of CTNF-type OA1 is also slightly higher for large businesses with the difference being statistically insignificant. Small and micro businesses are given more CTRS-type OA1s on average (approximately 9 percentage points), on the other hand. 

The cross table for the OA2 type, conditional on the event that the OA1
type is CTNF, is given in Table \ref{tbl:xtable.oa2}. The CTEQ and CTRS
types are infrequent as an OA2 type among those that have a CTNF-type
OA1. It is notable that small businesses show considerably higher
probabilities of receiving NOA-type OA2s than large ones. I will attempt
to identify using the DID methodology whether the effect is associated
with the AIA intervention, with other sources of change over time
controlled for.

\begin{table}[H]
  \caption{Cross table for OA2 types for applications with CTNF-type OA1}
  \label{tbl:xtable.oa2}
  \begin{center}
    \begin{tabular}{|*{7}{c|}}
      \hline
      Type & NOA & CTEQ & CTRS & CTNF & CTFR & Sum\\ \hline
      \multirow{2}{*}{Large} & 406 & 1 & 5 & 66 & 302 & 780\\
      & (52.1) & (0.1) & (0.6) & (8.5) & (38.7) & (100.0)\\ \hline
      \multirow{2}{*}{Small} & 123 & 1 & 3 & 11 & 56 & 194\\
      & (63.4) & (0.5) & (1.5) & (5.7) & (28.9) & (100.0)\\ \hline
      \multirow{2}{*}{Micro} & 10 & 0 & 1 & 1 & 5 & 17\\
      & (58.8) & (0.0) & (5.9) & (5.9) & (29.4) & (100.0)\\ \hline
      \multirow{2}{*}{Sum} & 539 & 2 & 9 & 78 & 363 & 991\\
      & (54.4) & (0.2) & (0.9) & (7.9) & (36.6) & (100.0)\\ \hline
    \end{tabular}
  \end{center}
  \footnotesize
  \emph{Note:} Row-wise percentage shares are in parentheses.
\end{table}

\begin{figure}[H]
  \caption{Delay until receipt of CTNF-type OA1}
  \label{fig:wait1}
  \begin{center}
    \includegraphics[width=.9\columnwidth]{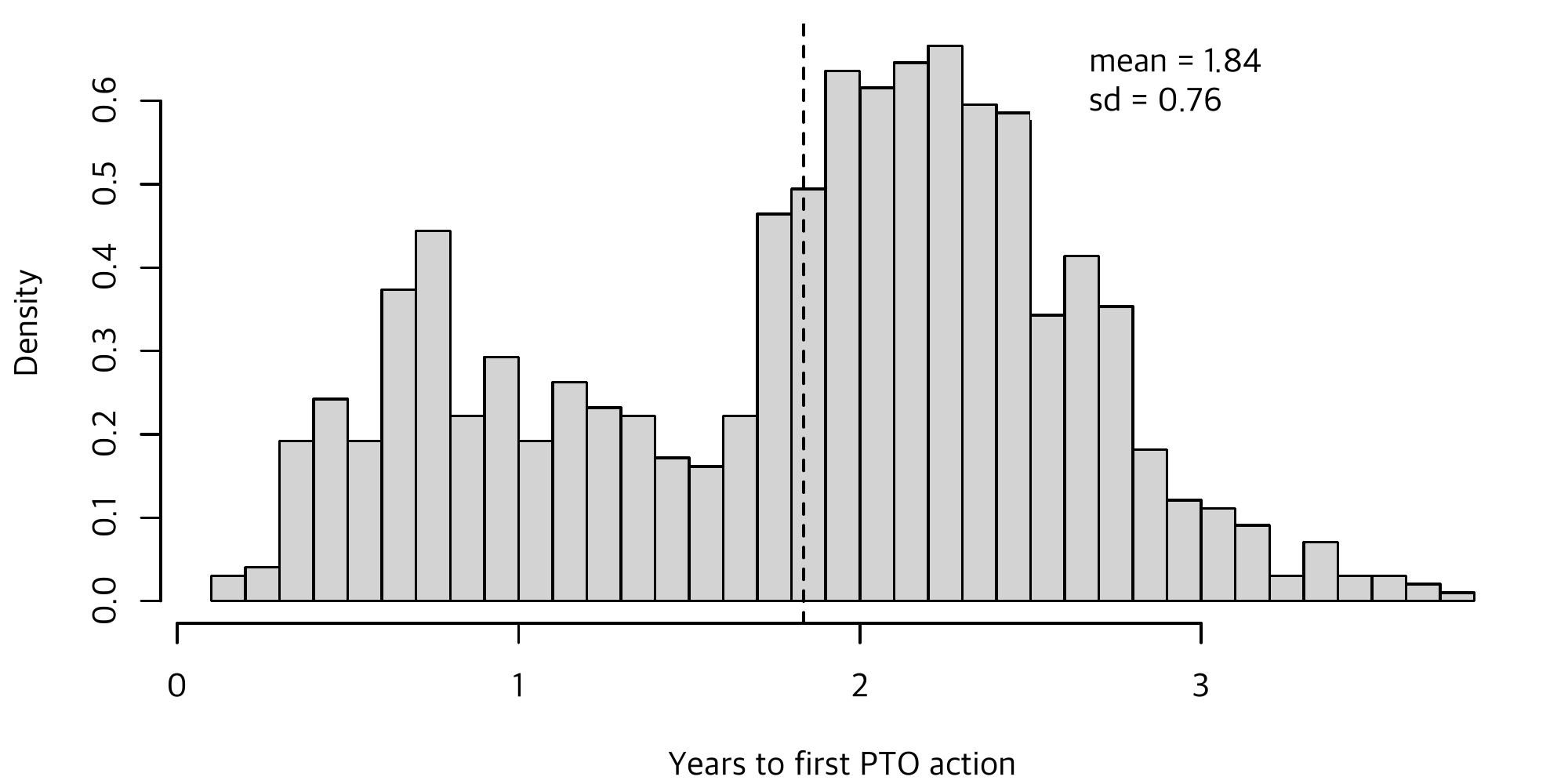}
  \end{center}
\end{figure}

The quality of the application as filed and the quality of the
applicant’s response to the first Non-Final Rejection are important
determinants in the resulting type of OA2. To control for
them, I note that the quality of the application as filed can be
correlated with the complexity of the initial Non-Final Rejection and
also with the difficulty to overcome that rejection. That difficulty can
influence the strength of the applicant’s response, which directly
affects the type of OA2 that is received. Based on this observation,
instead of directly processing the applications as filed, I consider the
delay between the application filing date and the date of receipt of
CTNF-type OA1 as a possible measure of the quality of the application as
filed. Delayed receipt of a CTNF-type OA1 may indicate that the
application as filed contains more flaws, although I recognize that the
delay may also be influenced by the flux of new patent applications at
the USPTO. The delay until receipt of a CTNF-type OA1 since filing is
distributed as in Figure \ref{fig:wait1}. On average, receipt of a
CTNF-type OA1 took 1.84 years (approximately 1 year and 10 months) since
filing, and the standard deviation is 0.76 years (approximately 9
months).

\begin{figure}[H]
  \caption{Delay until applicant response to CTNF-type OA1}
  \label{fig:adelay}
  \begin{center}
    \includegraphics[width=.9\columnwidth]{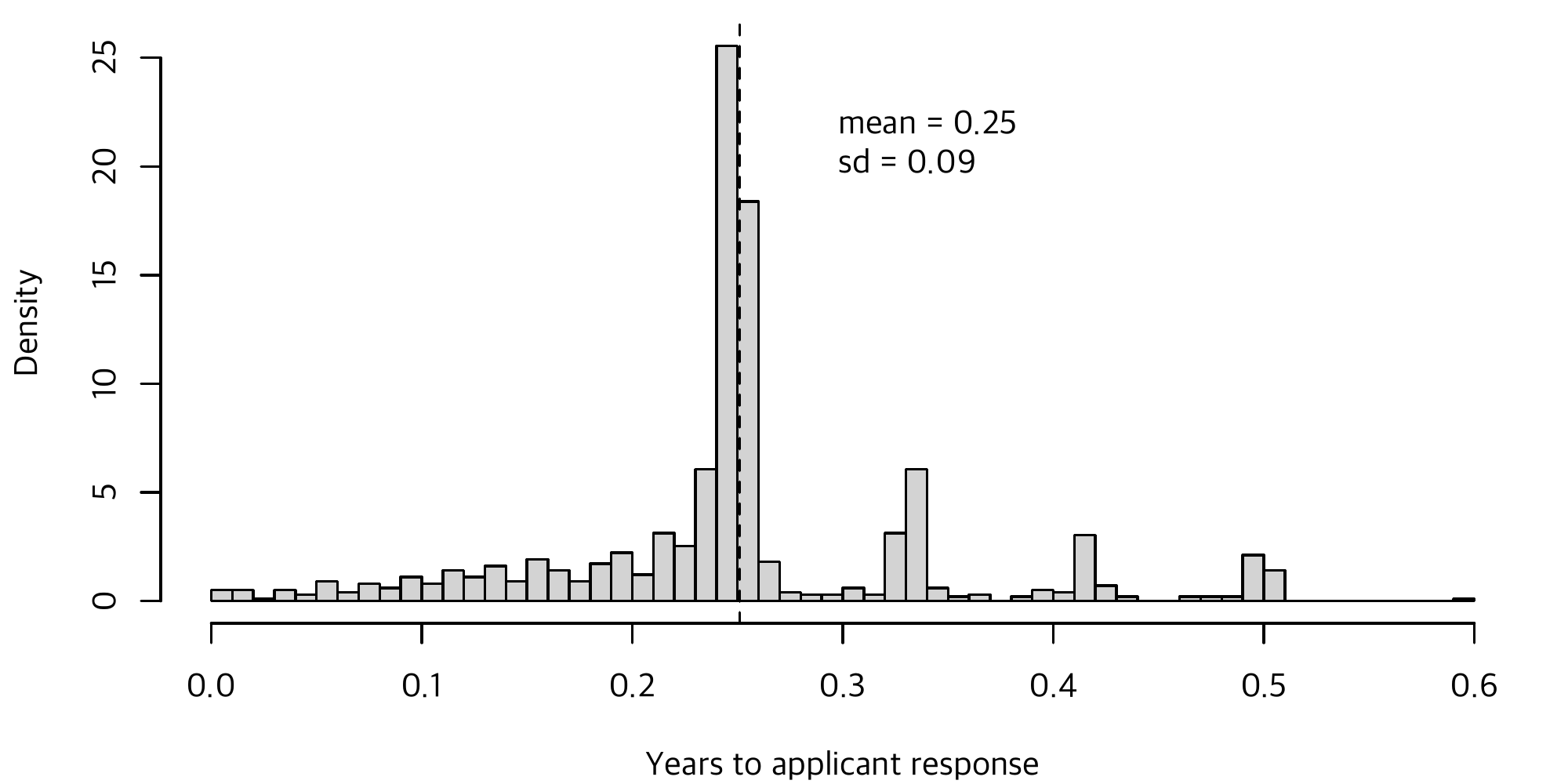}
  \end{center}
\end{figure}

The delay of the applicant’s response since receipt of a CTNF-type OA1 would be related with the difficulty in addressing the issues raised by USPTO. Except one application that took 2.6 years for a response, the distribution of this delay is given in Figure \ref{fig:adelay}. The mean length of delay is approximately 3 months with a standard deviation of approximately 1 month.

\begin{figure}[H]
  \caption{Delay until receipt of OA2 since applicant response to CTNF-type OA1}
  \label{fig:wait2}
  \begin{center}
    \includegraphics[width=.9\columnwidth]{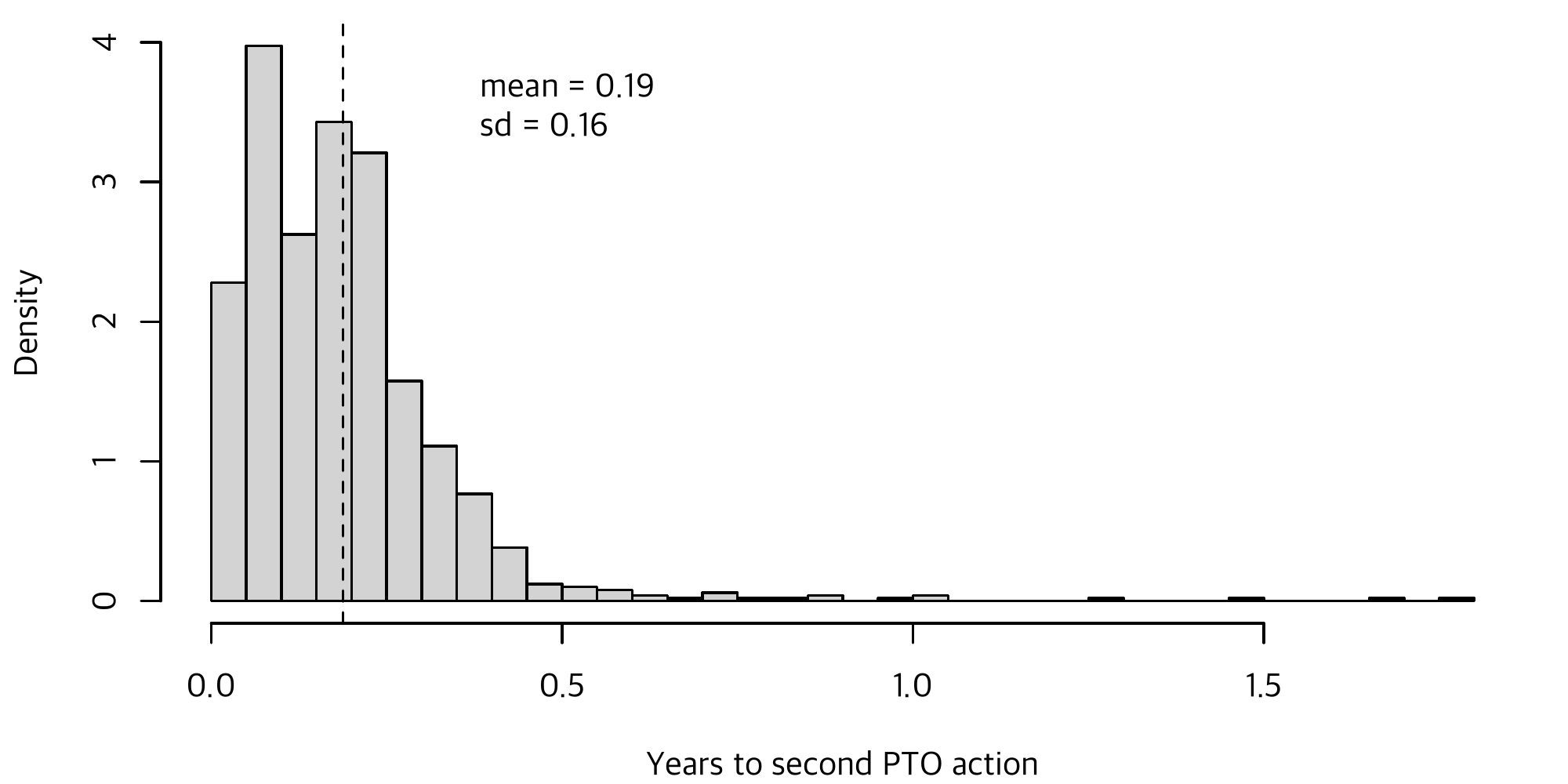}
  \end{center}
\end{figure}

Similarly, the delay until the receipt of an OA2 from the applicant response date may be correlated with the quality of the applicant’s response. This delay is distributed as in Figure \ref{fig:wait2}.

The number of rejected claims in OA1 is a more direct indicator of the
difficulty of addressing the issues raised by the USPTO. Non-Final Rejections
provide information on the sections in 35 U.S.C. under which
claims are rejected. These sections are
\textsection 101, \textsection 102(a), \textsection 102(b), \textsection
102(e), \textsection 103(a), \textsection 103(e), and \textsection 112
for the collected data. Table \ref{tbl:usc} summarizes the distributions
of the claim numbers rejected under these 35 U.S.C. sections, which is
used in the analyses in Section \ref{sec:analysis}.


\begin{table}[H]
  \caption{Distributions of claims rejected under 35 U.S.C. sections}
  \label{tbl:usc}
  \begin{center}
    \begin{tabular}{|*{8}{c|}}
      \hline
      Number of & \multicolumn{7}{c|}{35 USC \textsection}\\ \cline{2-8}
      rejected claims & 101 & 102(a) & 102(b) & 102(e) & 103(a) & 103(e) & 112\\
      \hline
      0 & 858 & 962 & 579 & 857 & 331 & 990 & 672\\ 
      1--5 & 51 & 8 & 130 & 33 & 189 & 0 & 153\\ 
      6--10 & 35 & 11 & 102 & 36 & 155 & 0 & 61\\ 
      11--15 & 24 & 4 & 78 & 24 & 100 & 0 & 32\\ 
      16--20 & 16 & 2 & 50 & 20 & 127 & 0 & 36\\ 
      21--25 & 4 & 3 & 25 & 8 & 39 & 0 & 18\\ 
      26--30 & 2 & 0 & 10 & 5 & 23 & 0 & 6\\ 
      31-- & 1 & 1 & 17 & 8 & 27 & 1 & 13\\ \hline
      Sum & 991 & 991 & 991 & 991 & 991 & 991 & 991\\
      \hline
    \end{tabular}
  \end{center}
\end{table}

\begin{figure}[H]
  \caption{Cumulative distribution of OA2 receipt date}
  \label{fig:cdf}
  \begin{center}
    \includegraphics[width=.9\columnwidth]{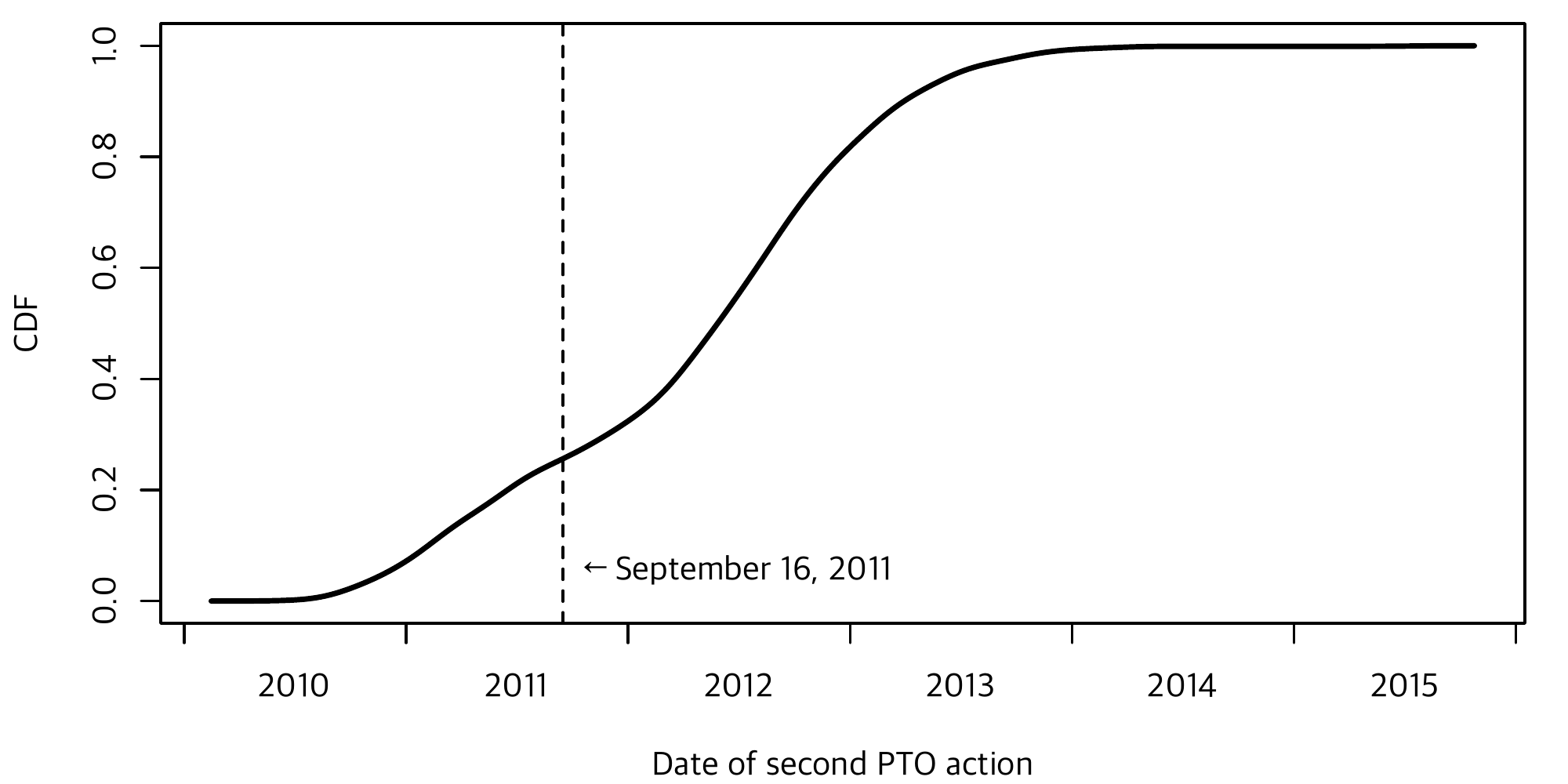}
  \end{center}
\end{figure}

One of the key variables in the present research is whether OA2 was received before or after the enactment of the AIA. Limited to the applications that have a CTNF-type OA1, the estimated cumulative distribution function (CDF) of the date of receiving OA2 is given in Figure \ref{fig:cdf}. A point on the estimated CDF represents the proportion of considered applications that receive an OA2 by that date. For example, 22.9\% of the applications have received an OA2 by September 16, 2011 (depicted by the vertical dashed line).

\begin{figure}[H]
  \caption{Distribution of the number of independent claims for applications with CTNF-type OA1}
  \label{fig:indclaims}
  \begin{center}
    \includegraphics[width=.9\columnwidth]{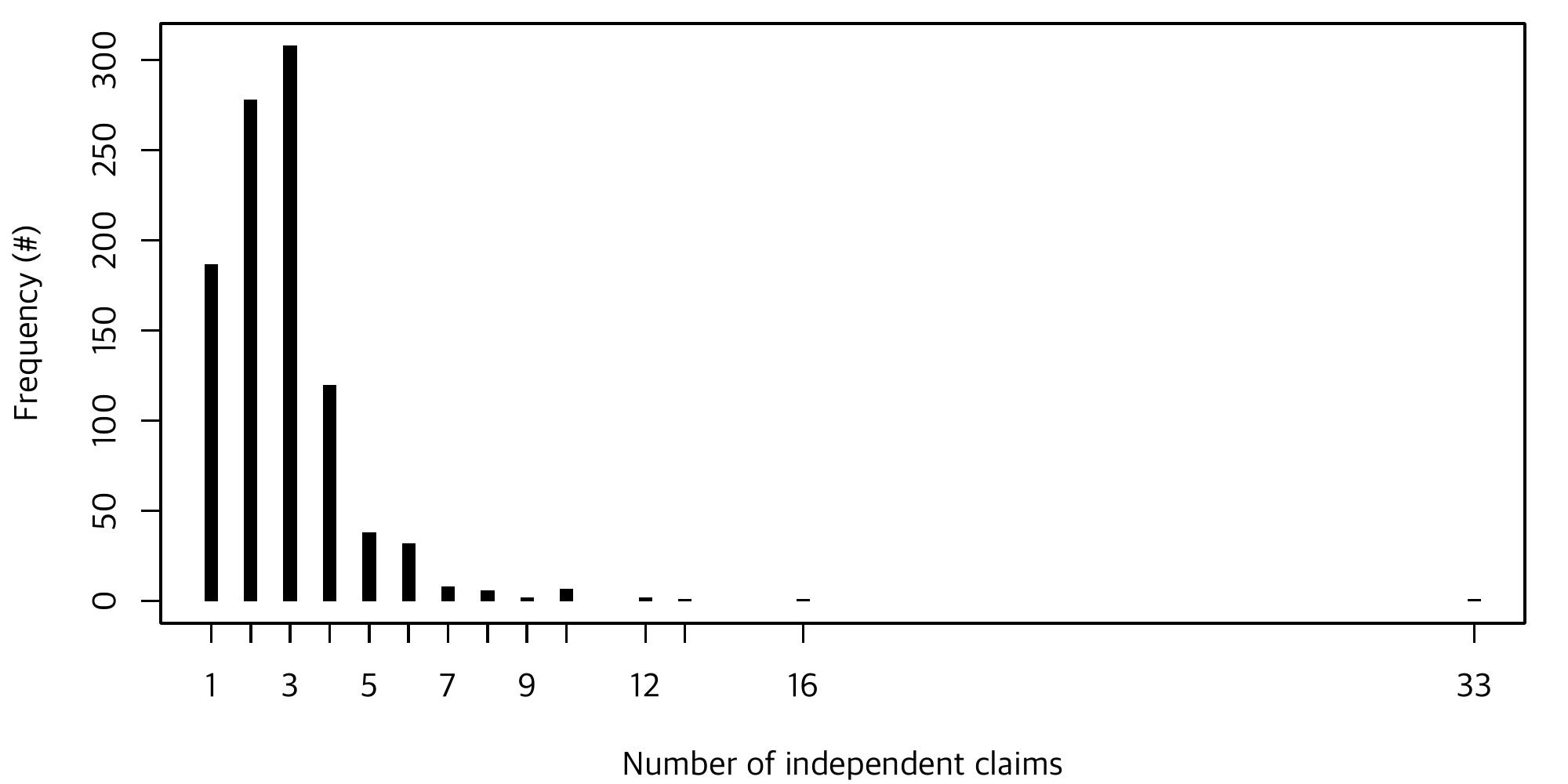}
  \end{center}
\end{figure}

The number of independent claims, described in Figure \ref{fig:indclaims}, can also be indicative of the complexity of the patent application. Most applications (97.2\%) have six or less independent claims.

Finally, I control for the total number of parent applications, the number of patented parent applications, the number of parent applications for each continuity type and the earliest priority date. These data points attempt to capture examination on matter related to the current application prior to its filing.

The dependent variable for the following analyses is the indicator for a successful response to a CTNF-type OA1, i.e., the dependent variable ‘success’ takes on the value 1 if the OA2 type is not CTNF or CTFR. This dependent variable and the independent variables to be used in this work are listed in Table \ref{tbl:varlist} below.

\begingroup \def\0{}
\begin{longtable}{|c|l|}
  \caption{List of control variables}
  \label{tbl:varlist}
  \\
  \hline
  Variable Name & Meaning\\
  \hline
  \endfirsthead
  \hline
  Variable Name & Meaning\\
  \endhead
  \multicolumn{2}{r}{(Continued on next page)}
  \endfoot
  \endlastfoot
  \0success & =1 if OA2 type is not CTFR or CTNF\\ \hline
  \0SmallMicro & =1 if small or micro business\\ \hline
  \0AIA & =1 if OA2 is received in the post-intervention period\\ \hline
  \0numIndClaims & number of independent claims\\ \hline
  \0dblPat & =1 if CTNF-type OA1 contains double patenting\\ \hline
  \0wait1 & delay from filing to receipt of OA1 (in years)\\ \hline
  \0adelay & delay from receipt of OA1 to applicant response (in years)\\ \hline
  \0wait2 & delay from applicant response to receipt of OA2 (in years)\\ \hline
  \0US & =1 if the corresponding address is in the United States\\ \hline
  \0numParents & number of parent applications\\ \hline
  \0noparent & =1 if number of parent applications is zero\\ \hline
  \0pryears & length of period (in years) between priority date and filing date;\\
  & 0 if there are no parent applications.\\ \hline
  \0numParentTypeCON & number of parent applications with CON type continuity\\ \hline
  \0numParentTypeCIP & number of parent applications with CIP type continuity\\ \hline
  \0numParentTypeDIV & number of parent applications with DIV type continuity\\ \hline
  \0numParentTypeNST & number of parent applications with NST type continuity\\ \hline
  \0numParentTypePRO & number of parent applications with PRO type continuity\\ \hline
  \0USC101 & number of claims rejected under 35 U.S.C. § 101\\ \hline
  \0USC102a & number of claims rejected under 35 U.S.C. § 102(a)\\ \hline
  \0USC102b & number of claims rejected under 35 U.S.C. § 102(b)\\ \hline
  \0USC102e & number of claims rejected under 35 U.S.C. § 102(e)\\ \hline
  \0USC103a & number of claims rejected under 35 U.S.C. § 103(a)\\ \hline
  \0USC103e & number of claims rejected under 35 U.S.C. § 103(e)\\ \hline
  \0USC112 & number of claims rejected under 35 U.S.C. § 112\\ \hline
  \0hasUSC101 & =1 if USC101 $> 0$\\ \hline
  \0hasUSC102a & =1 if USC102a $> 0$\\ \hline
  \0hasUSC102b & =1 if USC102b $> 0$\\ \hline
  \0hasUSC102e & =1 if USC102e $> 0$\\ \hline
  \0hasUSC103a & =1 if USC103a $> 0$\\ \hline
  \0hasUSC103e & =1 if USC103e $> 0$\\ \hline
  \0hasUSC112 & =1 if USC112 $> 0$\\ \hline
\end{longtable}
\endgroup

Summary statistics for the dependent variable and the variables in Table \ref{tbl:varlist} are provided in Table \ref{tbl:sum}.

\begingroup \def\0{}
\begin{longtable}{|*{8}{c|}}
  \caption{Summary statistics}
  \label{tbl:sum}
  \\
  \hline
  Variable & Obs & Mean & SD & Min & Median & Max & \# of 1's\\
  \hline
  \endfirsthead
  \hline
  Variable & Obs & Mean & SD & Min & Median & Max & \# of 1's\\
  \hline
  \endhead
  \hline
  \multicolumn{8}{r}{(Continued on next page)}
  \endfoot
  \hline
  \endlastfoot
  \0success & 991 & 0.553 & \phantom{0}0.497 & 0 & 1 & 1 & 548\\
  \0SmallMicro & 991 & 0.213 & \phantom{0}0.410 & 0 & 0 & 1 & 211\\
  \0AIA & 991 & 0.743 & \phantom{0}0.437 & 0 & 1 & 1 & 736\\
  \0numIndClaims & 991 & 2.832 & \phantom{0}1.902 & 1 & 3 & 33 & \\
  \0dblPat & 991 & 0.201 & \phantom{0}0.401 & 0 & 0 & 1 & 199\\
  \0wait1 & 991 & 1.835 & \phantom{0}0.760 & 0.175 & 1.984 & 3.707 & \\
  \0adelay & 991 & 0.253 & \phantom{0}0.119 & 0 & 0.249 & 2.647 & \\
  \0wait2 & 991 & 0.188 & \phantom{0}0.158 & 0 & 0.173 & 1.770 & \\
  \0US & 991 & 0.978 & \phantom{0}0.147 & 0 & 1 & 1 & 969\\
  \0numParents & 991 & 0.932 & \phantom{0}1.591 & 0 & 0 & 26 & \\
  \0noparent & 991 & 0.506 & \phantom{0}0.500 & 0 & 1 & 1 & 501\\
  \0pryears & 991 & 1.752 & \phantom{0}2.690 & 0 & 0 & 14.740 & \\
  \0numParentTypeCON & 991 & 0.297 & \phantom{0}0.667 & 0 & 0 & 5 & \\
  \0numParentTypeCIP & 991 & 0.118 & \phantom{0}0.504 & 0 & 0 & 10 & \\
  \0numParentTypeDIV & 991 & 0.112 & \phantom{0}0.393 & 0 & 0 & 5 & \\
  \0numParentTypeNST & 991 & 0.025 & \phantom{0}0.157 & 0 & 0 & 1 & \\
  \0USC101 & 991 & 1.181 & \phantom{0}3.974 & 0 & 0 & 37 & \\
  \0USC102a & 991 & 0.308 & \phantom{0}2.215 & 0 & 0 & 35 & \\
  \0USC102b & 991 & 4.800 & \phantom{0}9.562 & 0 & 0 & 160 & \\
  \0USC102e & 991 & 1.772 & \phantom{0}6.106 & 0 & 0 & 65 & \\
  \0USC103a & 991 & 8.286 & 10.741 & 0 & 5 & 137 & \\
  \0USC103e & 991 & 0.031 & \phantom{0}0.985 & 0 & 0 & 31 & \\
  \0USC112 & 991 & 3.011 & \phantom{0}6.990 & 0 & 0 & 60 & \\
  \0hasUSC101 & 991 & 0.134 & \phantom{0}0.341 & 0 & 0 & 1 & 133\\
  \0hasUSC102a & 991 & 0.029 & \phantom{0}0.169 & 0 & 0 & 1 & 29\\
  \0hasUSC102b & 991 & 0.416 & \phantom{0}0.493 & 0 & 0 & 1 & 412\\
  \0hasUSC102e & 991 & 0.135 & \phantom{0}0.342 & 0 & 0 & 1 & 134\\
  \0hasUSC103a & 991 & 0.666 & \phantom{0}0.472 & 0 & 1 & 1 & 660\\
  \0hasUSC112 & 991 & 0.322 & \phantom{0}0.467 & 0 & 0 & 1 & 319\\
  \0hasUSC103e & 991 & 0.001 & \phantom{0}0.032 & 0 & 0 & 1 & 1\\
\end{longtable}

\noindent
\footnotesize
\emph{Note:} ``\# of 1's'' is the number of ones in the sample of dummy variables.

\endgroup

\section{Analysis}
\label{sec:analysis}

\subsection{Methodology}

I use the difference-in-differences (DID) method to measure the effect of the AIA on the success rate, the rate of successful response to initial Non-Final Rejection, of small and large businesses. The DID method is widely used in social sciences to measure causal effects of policy interventions. (See Card and Krueger, 1994, for example.) In general terms, the DID method is used in the following manner.

There exists a sample consisting of two groups of observational units. One group is the treatment group and the other the control group. Given a chosen intervention date, the periods are dichotomized into a pre-intervention and a post-intervention period. The DID method begins with measuring the average values of the dependent variable for each group in each period. Table \ref{tbl:did0} exhibits these averages as A00, A01, A10, and A11, respectively, where A00 and A01 are the averages for the control group in the pre-intervention and post-intervention periods, respectively, and A10 and A11 repeat the same practice for the treatment group.

\begin{table}[H]
  \caption{Diagram for difference-in-differences}
  \label{tbl:did0}
  \begin{center}
    \begin{tabular}{|c|c|c|c|}
      \hline
      & Control group & Treatment group & Difference\\ \hline
      Pre-intervention & A00 & A10 & A10 – A00\\ \hline
      Post-intervention & A01 & A11 & A11 – A01\\ \hline
      Change & A01 – A00 & A11 – A10 & (A11 – A10) – (A01 – A00)\\ \hline
    \end{tabular}
  \end{center}
\end{table}

The change in the average dependent variable for the treatment group (A11 – A10) consists of the treatment effect and the associated uncontrolled trend between the dichotomized periods:
\begin{equation}
  \label{eq:1}
  \text{A11}-\text{A10} = (\text{Treatment effect}) + (\text{Trend for treatment group}),
\end{equation}
where random errors are ignored due to the Law of Large Numbers. It is notable that the change $\text{A11}-\text{A10}$ contains not only the treatment effect of interest, but also the changes due to factors unrelated to the policy intervention. To find the treatment effect using (\ref{eq:1}), it is necessary to know the value of $\text{A11}-\text{A10}$ and the trend effect for the treatment group. The value of A11 – A10 can be estimated by the sample. To find the trend effect for the treatment group, we turn to the over-time change for the control group.

The over-time change for the control group consists of only the trend effect
\begin{equation}
  \label{eq:2}
  \text{A01}-\text{A00} = (\text{Trend for control group})
\end{equation}
because the control group is not treated. The random errors are again ignored due to the Law of Large Numbers. Here the key assumption for the DID method is implemented, that is, the parallel trends assumption which states that the treatment and control groups both share the same trend effect (Wing et al., 2018, Schiozer et al., 2021). Under this assumption, the treatment effect is can be found by taking the difference between (\ref{eq:1}) and (\ref{eq:2}) -- the difference-in-differences (DID):
\begin{equation}
  \label{eq:3}
  \text{DID} = (\text{A11}- \text{A10}) - (\text{A01}-\text{A00}).
\end{equation}
The DID method is alternatively called a comparative interrupted time series design or a nonequivalent control group pretest design (Wing et al., 2018).

The DID method can be implemented by the standard ordinary least squares (OLS) regression. To estimate the DID effect, one can regress the dependent variable (Y) on the dummy variable (TR) for treatment group, another dummy variable (POST) for the post-treatment period, and the interaction of the two, that is, TR*POST. Then the DID estimate is simply the coefficient on the interaction term (Wing et al., 2018, p. 456). The standard errors and associated confidence intervals to use for inferences are reported by standard statistical software such as R.

The parallel trends assumption is vital for the regression of Y on TR, POST, and TR*POST to give a valid policy effect estimator. If this assumption is violated, the trend heterogeneity confounds the causal effect. In that case, one can control for the factors that are responsible for the trend heterogeneity. For example, if the distribution of observations over time for the treatment group differs from that of the control group, and there are other interventions within the vicinity of intervention date in question, year dummies or year-quarter dummies can be included to mitigate the effect of those other unwanted interventions. Also, quality measures can be controlled for if quality difference drives group-wise heterogeneity in trends. 

In these analyses, the treatment group is composed of the patent applications owned by small-business applicants (including micro-business applicants), and the control group is composed of those owned by large-business applicants. The date of the intervention in question is the date the AIA has been signed into law, September 16, 2011.

A complication arises with regard to Equation (\ref{eq:2}). The standard DID method states that the control group is free from policy intervention. However, in the case of this analysis, patent applications owned by large businesses may be affected by the AIA just as those of small businesses are. Because of this, I look at the ‘relative policy effect’, defined as the difference in the AIA’s effects on small business patent applications and its effects on large business patent applications. It can be observed that if the parallel tends assumption is satisfied, the DID approach will consistently estimate the ‘relative policy effect’. To demonstrate this, see that Equation (1) can be expressed as
\begin{equation}
  \label{eq:1a}
  \text{A11}-\text{A10} = (\text{Policy effect for treatment group})+ (\text{Trend for treatment group}),
\end{equation}
while Equation (2) can be generalized to
\begin{equation}
  \label{eq:2a}
  \text{A01}-\text{A00} = (\text{Policy effect for control group})+( \text{Trend for control group}).
\end{equation}
The `relative policy effect' is defined as the difference of the two policy effects in (\ref{eq:1a}) and (\ref{eq:2a}). Under the assumption that the trends are identical for the treatment group and the control group, the DID statistic estimates the ‘relative policy effect’, which is the difference between the policy effects between the two groups. In other words, the relative policy effect is the heterogeneity in the policy effects between the two groups. A positive DID statistic indicates that small-businesses applicants enjoy a relatively higher success rate than large-business applicants after the AIA became effective. This could suggest that small businesses received a more favorable treatment within the USPTO post-AIA in comparison to large businesses. A negative DID statistic indicates otherwise.

\subsection{Results}

\begin{figure}[H]
  \caption{Success ratios during centered three-month periods}
  \label{fig:ym3}
  \begin{center}
    \includegraphics[width=.9\columnwidth]{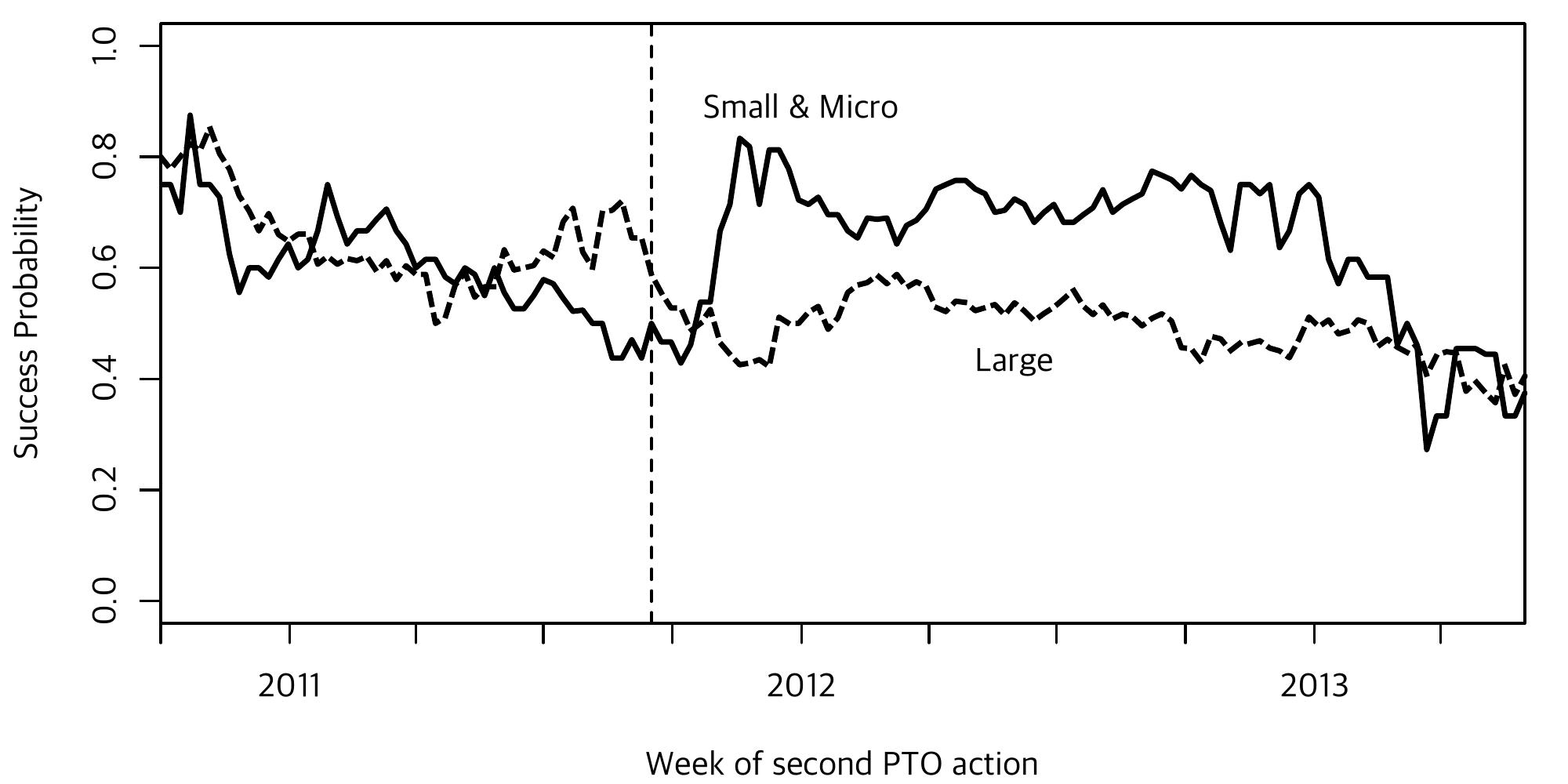}
  \end{center}

  \footnotesize
  \emph{Note:} The vertical dashed line is for September 16, 2011.
\end{figure}

Prior to conducting formal regression analysis, I present trends in success probability for patent applications owned by large-business applicants and those owned by small-businesses applicants. Figure \ref{fig:ym3} exhibits centered three-month moving averages of successful responses for each of large and small business groups in the sample of 991 applications, where ‘success’ is defined in Table \ref{tbl:varlist}, and the centered moving average at a certain date is defined as the ratio of successes in the 91-day period centered at that date. For example, the moving average for December 1, 2011 is the ratio of successes from October 17, 2011 (45 days prior) to January 15, 2012 (45 days posterior).

Prior to the AIA, the patent applications owned by large-business applicants and those owned by small-business applicants do not seem to display a systematic difference between their success rates. However, for more than a year starting approximately one month after the introduction of AIA, small-business-owned patent applications exhibit systematically higher success rates in comparison to large-business-owned patent applications. Loosely speaking, regression analyses in this section measure the average change of the success-rate differentials over time between small and large businesses.

Results from the formal DID regressions are presented in Table \ref{tbl:regmain}.

\begingroup \def\0{\em}
\begin{longtable}{@{}l*{4}{@{~~~~}r@{}l@{}}}
  \caption{Difference-in-differences estimation results}
  \label{tbl:regmain}
  \\
  \hline
  success & \multicolumn{2}{c}{(1)} & \multicolumn{2}{c}{(2)} & \multicolumn{2}{c}{(3)} & \multicolumn{2}{c}{(4)}\\
  \hline
  \endfirsthead
  \hline
  success & \multicolumn{2}{c}{(1)} & \multicolumn{2}{c}{(2)} & \multicolumn{2}{c}{(3)} & \multicolumn{2}{c}{(4)}\\
  \hline
  \endhead
  \hline
  \multicolumn{8}{r}{(Continued on next page)}
  \endfoot
  \endlastfoot
  SmallMicro*AIA & 0.2347 & *** & 0.2037 & ** & 0.1850 & ** & 0.1655 & **\\ 
  SmallMicro & -0.0502 &  & -0.0310 &  & -0.0100 &  & -0.0035 &\\ 
  AIA & -0.1550 & *** & -0.2385 &  & -0.1813 &  & -0.1508 &\\ 
  log(numIndClaims) &  &  &  &  & -0.0806 & *** & -0.0621 & **\\ 
  dblPat &  &  &  &  & 0.0244 &  & -0.0227 &\\ 
  wait1 &  &  &  &  & -0.0210 &  & 0.0092 &\\ 
  adelay &  &  &  &  & -0.6348 & ** & -0.5031 & *\\ 
  wait2 &  &  &  &  & -0.6606 & *** & -0.5864 & **\\ 
  US &  &  &  &  & 0.0697 &  & 0.0967 &\\ 
  numParents &  &  &  &  & 0.0263 &  & 0.0413 & **\\ 
  noparent &  &  &  &  & 0.0310 &  & 0.0041 &\\ 
  pryears &  &  &  &  & -0.0218 & * & -0.0252 & **\\ 
  numParentTypeCON &  &  &  &  & 0.0471 &  & 0.0090 &\\ 
  numParentTypeCIP &  &  &  &  & -0.0628 &  & -0.0818 & *\\ 
  numParentTypeDIV &  &  &  &  & 0.0799 &  & 0.0281 &\\ 
  numParentTypeNST &  &  &  &  & -0.0198 &  & -0.0197 &\\ 
  USC101 &  &  &  &  &  &  & 0.0031 &\\ 
  USC102a &  &  &  &  &  &  & -0.0240 & **\\ 
  USC102b &  &  &  &  &  &  & -0.0038 & *\\ 
  USC102e &  &  &  &  &  &  & -0.0067 & *\\ 
  USC103a &  &  &  &  &  &  & -0.0049 & ***\\ 
  USC103e &  &  &  &  &  &  & -0.0309 & **\\ 
  USC112 &  &  &  &  &  &  & 0.0010 &\\ 
  hasUSC101 &  &  &  &  &  &  & -0.0564 &\\ 
  hasUSC102a &  &  &  &  &  &  & 0.1005 &\\ 
  hasUSC102b &  &  &  &  &  &  & -0.0577 &\\ 
  hasUSC102e &  &  &  &  &  &  & -0.0030 &\\ 
  hasUSC103a &  &  &  &  &  &  & -0.1290 & ***\\ 
  hasUSC112 &  &  &  &  &  &  & -0.0654 &\\ \hline
  Year-quarter dummies & No &  & Yes &  & Yes &  & Yes &\\ \hline
  Intercept & 0.6440 & *** & 0.0310 &  & 0.3507 &  & 0.4278 &\\ \hline
  $n$ & 991 &  & 991 &  & 991 &  & 991 &\\
  R-squared & 0.0255 &  & 0.0431 &  & 0.1080 &  & 0.1584 &\\
  Adj. R-squared & 0.0225 &  & 0.0234 &  & 0.0773 &  & 0.1174 &\\ \hline
\end{longtable}

\noindent
\footnotesize
\emph{Note:} Robust standard errors are in parentheses. The DID estimates are on the first row. ***, ** and * stand for statistical significance at the 1\%, 5\% and 10\% levels, respectively.

\endgroup

\bigskip

In Column (1) of Table \ref{tbl:regmain}, the intercept 0.6440 measures the estimated success rate for the control group (large businesses) during the pre-AIA period. The coefficient on the ‘SmallMicro’ variable is the average differential of the success rate between the treatment group (small businesses) and the control group (large businesses) in the pre-treatment period (pre-AIA). The estimated value -0.0502 means that the success rate of the treatment group is estimated to be lower than that of the control group in the pre-treatment period.

The coefficient on the AIA variable measures the average ‘before/after’ change in the success rate for the control group. Patent applications filed by large businesses experienced a decreased success rate (-0.1550). Though not shown in the table, the treatment group, on the contrary, experienced an increase in the success rate by 7.97 percentage points. In comparison to the control group’s experience of a decrease by 16.01 percentage points, the treatment group experienced a relative increase in success rate by 23.47 percentage points, which is the estimated coefficient (DID) on the interaction term. This effect is practically large in magnitude amounting to 36\% of large businesses' average success rate in the pre-AIA period, and is statistically significant. These results are summarized in Table \ref{tbl:did.simple}.

\begin{table}[H]
  \caption{Simple DID table}
  \label{tbl:did.simple}
  \begin{center}\def\0{\phantom{-}}%
    \begin{tabular}{|c|c|c|c|}
      \hline
      & Large & Small \& Micro & Difference\\ \hline
      Pre-invention & \00.6440 & 0.5938 & -0.0502\\ \hline
      Post-invention & \00.4890 & 0.6735 & \00.1845\\ \hline
      Change & -0.1550 & 0.0797 & \00.2347\\ \hline
    \end{tabular}
  \end{center}
\end{table}

\begin{figure}[H]
  \caption{Difference-in-differences graphically illustrated}
  \label{fig:did2x2}
  \begin{center}
    \includegraphics[width=.9\columnwidth]{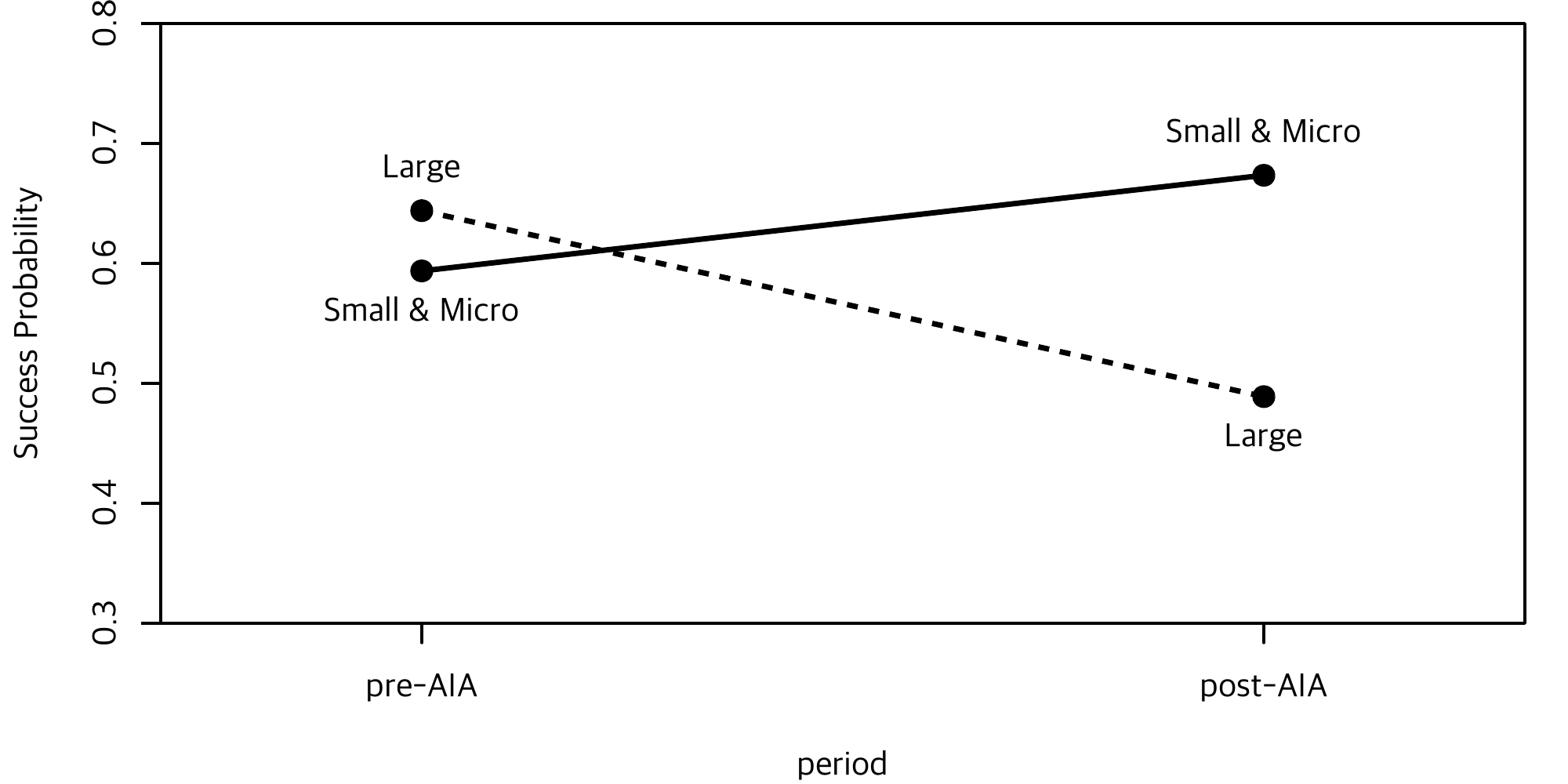}
  \end{center}
\end{figure}

A graphical illustration of Figure \ref{fig:did2x2} plots success rates for both groups during the pre- and post-AIA periods. In the pre-AIA period, applications owned by large businesses showed a slightly higher average success rate in comparison to those owned by small businesses by 5.02 percentage points according to Table \ref{tbl:did.simple}. This is in contrast with the post-AIA period. While the average success rate for applications owned by large businesses experienced a fall of success rate by 15.5 percentage points, the average success rate of those owned by small businesses rose by approximately 8 percentage points. The difference of 23.47 percentage points of the two changes is the estimated DID effect, i.e., the ‘relative policy effect’.

Column (2) controls for the year-quarter effects by including dummy variables (where ‘year-quarter’ means the year-quarter of the OA2 receipt date). These year-quarter dummy variables account for the changes in possible differences in the outflow of Office Actions in each quarter between the treatment and the control groups. As a result, the estimated effect slightly decreases in magnitude in comparison to Column (1) but remains statistically significant at the 5\% level.

Column (3) controls for various characteristics of the patent applications together with the year-quarter effects. These extra control variables are the log number of independent claims (`numIndClaims'), whether the CTNF-type OA1 contains remarks on double patenting (`dblPat'), the delay between filing and the receipt date of the OA1 (`wait1'), the delay between the OA1 receipt date and the applicant’s response date (`adelay'), the delay between the applicant’s response date and the receipt date of the OA2 (`wait2'), the country of corresponding address (`US'), the number of parent applications (`numParents'), the dummy variable indicator of no parent applications (`noparent'), the interval between the priority date and the filing date in years (`pryears'), the number of parent applications with continuity types CON, CIP, DIV and NST. The remaining continuity type PRO is omitted in order to avoid collinearity with ‘numParents’. The estimated DID effect decreases by 1.5 percentage points from Column (2), but the estimated value 0.1850 again seems considerable in magnitude and is statistically significant at the 5\% level.

Shifting focus to the other variables in Column (3), the log number of independent claims is negatively correlated with the success rate and its effect is statistically significant at the 1\% level. Double patenting does not have significant effects. The delay periods after the first Non-Final Rejection are estimated to have adverse effects on success rate. The ‘pryears’ variable is also negatively correlated with the success rate. It is notable that the inclusion of the ‘noparent’ dummy variable limits the comparison of the effects of ‘pryears’ to among patent applications that have a nonzero number of parent applications. The negative estimated coefficient (-0.0218), significant at the 10\% level, suggests that the longer the time difference between the priority date and the filing date, the lower the success rate among comparable patent applications.

Column (4) further controls for the numbers of claims rejected under each section within 35 U.S.C. The numbers of rejected claims under each section are negatively correlated with the success rate. This aligns with the intuition that there are more rejected claims in a Non-Final Rejection that is more difficult to overcome. The estimated DID effect again falls by approximately 2 percentage points, but is still large in magnitude (0.1655) and statistically significant.

I have defined the post-AIA period as September 16, 2011 until present day. It is interesting to observe the change or lack of change of the magnitude and statistical significance of the DID effect when the policy intervention date is changed. Specifically, I explored changing the intervention date to October, November, and December of 2011, respectively. These dates respectively mark the one month, two month, and three month date since the AIA was signed into law. For these intervention date alterations, the period between September 16, 2011 and the new intervention date was disregarded as a ‘gray period.’ The DID effects change to Columns (2)--(4) of Table \ref{tbl:periods} for model (4) in Table \ref{tbl:regmain}. The results remain robust.

\begin{table}[H]
  \caption{Various intervention periods}
  \label{tbl:periods}
  \begin{center}
    \begin{tabular}{@{}l*{4}{r@{}l}@{}}
      \hline
      success & \multicolumn{2}{c}{(1)} & \multicolumn{2}{c}{(2)} & \multicolumn{2}{c}{(3)} & \multicolumn{2}{c}{(4)}\\
      \hline
      SmallMicro*AIA & 0.1655 & ** & 0.1769 & ** & 0.1739 & ** & 0.1670 & **\\
      SmallMicro & -0.0035 &  & -0.0062 &  & -0.0080 &  & -0.0040 &\\
      AIA & -0.1508 &  & 2.1323 &  & 2.1955 &  & 2.2020 &\\
      \hline
      \multicolumn{9}{c}{Other variables omitted}\\
      \hline
      $n$  & 991 &  & 979 &  & 958 &  & 935 &\\
      R-squared  & 0.1584 &  & 0.1576 &  & 0.1558 &  & 0.1586 &\\
      Adj. R-squared  & 0.1174 &  & 0.1170 &  & 0.1142 &  & 0.1160 &\\
      \hline
    \end{tabular}
  \end{center}
  \footnotesize
  \emph{Note:} (1)=Model (4) in Table \ref{tbl:regmain}; (2)=comparison of pre 09/16/2011 and post 10/16/2011; (3) comparison of pre 09/16/2011 and post 11/16/2011; (4)=comparison of pre 09/16/2011 and post 12/16/2011. Robust standard errors are in parentheses. The DID estimates are on the first row. ***, ** and * stand for statistical significance at the 1\%, 5\% and 10\% levels, respectively.
\end{table}

\section{Concluding Remarks}
\label{sec:conclude}

This paper examines the effects of the AIA on small businesses that have
pending patent applications in the Search and Examination phase of
patent prosecution. Data was collected from the USPTO’s PatFT search
engine and Patent Center tool, and Google Patents. The results of this
empirical study show that after the enactment of the AIA on September
16, 2011, small-business applicants experienced a statistically
significant higher success rate of 16.5 to 23.5 percentage points,
measured by difference-in-differences (DID), than that of large-business
applicants. The estimated effect on small businesses is robust in that
the inclusion or removal of control variables in the DID regressions
does not considerably affect the DID coefficient and its statistical
significance.

While this study uses a large enough sample to show statistically
significant DID effects, collection of further data would allow for an
assessment of long-term effects on small-business applicants. This may
be desirable since the AIA was fully implemented over the course of 18
months (USPTO, 2011). Also, this study only uses patent applications
that were ultimately issued as a patent. Investigation of full data
including abandoned patent applications is left for future research
despite the data for such applications having limited public
accessibility.

\section*{References}

\begin{list}{}{\leftmargin=\parindent \itemindent=-\leftmargin}
\item Braun, R. G. (2012). America Invents Act: First-to-file and a race to the patent office, 8 \emph{Ohio St. Entrepren. Bus. L. J.}, 47.

\item Card, D., and A. B. Krueger (1994). Minimum wages and employment: A case study of the fast-food industry in New Jersey and Pennsylvania, \emph{American Economic Review}, 84, 772--793.

\item Case, J. (2013). How the America Invents Act hurts American
  inventors and weakens incentives to innovate, \emph{University of
    Missouri-Kansas City Law Review}, 82(1), 29--77.


\item Lerner, J., A. Speen, and A. Leamon (2015). The Leahy-Smith America Invents Act: A Preliminary Examination of Its Impact on Small Businesses, \emph{SBA Office of Advocacy}. Available at: https://www.sba.gov/sites/default/files/advocacy/rs429tot\_AIA\_Impact\_on\_SB.pdf.

\item Schiozer, R. F., F. A. Mourad, and T. C. Martins (2021). A
  tutorial on the use of differences-in-differences in management,
  finance, and accounting. \emph{Revista de Administração
    Contemporânea}, 25(1), e200067. Available at:
  https://doi.org/10.1590/1982-7849rac2021200067.

\item Sutton, P. J. (2014). AIA's impact upon small businesses,
  \emph{World Intellectual Property Review}. Available at:
  https://www.worldipreview.com/article/aia-s-impact-upon-small-businesses.

\item United States Patent and Trademark Office (2011). America Invents Act: Effective dates. Available at:
  https://www.uspto.gov/sites/default/files/aia\_implementation/aia-effective-dates.pdf.

\item Vandenburg, E. P. (2014). America Invents Act: How it affects small businesses, 50 \emph{Idaho Law Review}.201. Available at:
  https://digitalcommons.law.uidaho.edu/idaho-law-review/vol50/iss1/8.

\item Wing, C., K. Simon, and R. A. Bello-Gomez (2018). Designing difference in difference studies: Best practicies for public health policy research, \emph{Annual Reviews of Public Health}, 39, 453--469.
\end{list}

\end{document}